\documentclass[twocolumn,prl]{revtex4}
\usepackage{float}
\usepackage{amsfonts}
\usepackage{makeidx}
\usepackage{graphicx}
\usepackage{amsmath,amssymb,graphics,epsfig,color,times,bbm}

\begin{document}

\title{Local complementation rule for continuous-variable four-mode unweighted graph states}
\author{Jing Zhang$^{\dagger }$}
\affiliation{State Key Laboratory of Quantum Optics and Quantum
Optics Devices, Institute of Opto-Electronics, Shanxi University,
Taiyuan 030006, P.R.China}

\begin{abstract}
The local complementation rule is applied for continuous-variable
(CV) graph states in the paper, which is an elementary graph
transformation rule and successive application of which generates
the orbit of any graph states. The corresponding local Gaussian
transformations of local complementation for four-mode unweighted
graph states were found, which do not mirror the form of the local
Clifford unitary of qubit exactly. This work is an important step to
characterize the local Gaussian equivalence classes of CV graph
states.
\end{abstract}

\maketitle

Entanglement lies at the heart of quantum mechanics and plays a
crucial role in quantum information processing. Recently, special
types of multipartite entangled states, the so-called the graph
states \cite{one,two}, have moved into the center of interest. A
graph quantum state is described by a mathematical graph, i.e. a set
of vertices connected by edges. A vertex represents a physical
system, e. g. a qubit (2-dimensional Hilbert space), qudit
(d-dimensional Hilbert space), or CV (continuous Hilbert space). An
edge between two vertices represents the physical interaction
between the corresponding systems. An interesting feature is that
many entanglement properties of graph states are closely related to
their underlying graphs. They not only provide an efficient model to
study multiparticle entanglement \cite{one}, but also find
applications in quantum error correction \cite{three,four},
multi-party quantum communication \cite{five} and most prominently,
serve as the initial resource in one-way quantum computation
\cite{six}. Considerable efforts have been stepped toward generating
and characterizing cluster state with linear optics experimentally
\cite{seven,eight,nine,ten}. The principle feasibility of one-way
quantum computing model has been experimentally demonstrated through
photon cluster state successfully \cite{seven,ten}.

Most of the concepts of quantum information and computation have
been initially developed for discrete quantum variables, in
particular two-level or spin-$\frac{1}{2}$ quantum variables
(qubits). In parallel, quantum variables with a continuous spectrum,
such as the position and momentum of a particle or amplitude and
phase quadrature of an electromagnetic field, in informational or
computational processes have attracted a lot of interest and appears
to yield very promising perspectives concerning both experimental
realizations and general theoretical insights \cite{eleven,twelve},
due to relative simplicity and high efficiency in the generation,
manipulation, and detection of CV state. Although up to six-qubit
single-photon cluster states have been created via postselection
using nonlinear and linear optics, the deterministic, unconditional
realization of optical cluster states would be based on continuous
variables. CV cluster and graph states have been proposed
\cite{thirteen}, which can be generated by squeezed state and linear
optics \cite{thirteen1,thirteen2}, and demonstrated experimentally
for four-mode cluster state \cite{forteen}. The one-way CV quantum
computation was also proposed with CV cluster state \cite{fifteen}.
Moreover, the protocol of CV anyonic statistics implemented with CV
graph states is proposed \cite{fifteen1}.

One of the interesting issues on entanglement is how to define the
equivalence of two entangled states. The transformations of qubit
graph states under local Clifford operations were studied by Hein
\cite{one} and Van den Nest \cite{sixteen}. They translate the
action of local Clifford operations on qubit graph states into
transformations on their associated graphs, that is, to derive
transformations rules called the local complement rule, stated in
purely graph theoretical terms, which completely characterize the
evolution of graph states under local Clifford operations. The
corresponding local Clifford unitary is a single and simple form.
The successive application of this rule suffices to generate the
complete orbit of any qubit graph state under local Clifford
operations. In this paper, the local complement rule for CV
four-mode unweighted graph state is applied and the corresponding
local Clifford transformations (also called local Gaussian
transformation for CV) for four-mode graph state were found. The
local Gaussian equivalence classes of CV four-mode unweighted graph
states can be obtained by this way. It was shown that the
corresponding local Gaussian unitary can not exactly mirror that for
qubit, which is not a single form compared with qubit. This result
shows the complexity of CV quantum systems and stimulate the
research on the local Gaussian equivalence of CV graph states.
Although only focusing on the CV four-mode unweighted graph states,
this work makes an important step in the direction of addressing the
general question "What are the graph transformation rules that
describe local unitary equivalence of any CV graph states?".

The CV operations are reviewed firstly that follow the standard
prescription given in Ref.\cite{seventeen}. The Pauli $X$ and $Z$
operators of qubit are generalized to the Weyl-Heisenberg group,
which is the group of phase-space displacements. For CVs, this is a
Lie group with generators
$\hat{x}=(\hat{a}+\hat{a}^\dagger)/\sqrt{2}$ (quadrature-amplitude
or position) and $\hat{p}=-i(\hat{a}-\hat{a}^\dagger)/\sqrt{2}$
(quadrature-phase or momentum) of the electromagnetic field as the
CV system. These operators satisfy the canonical commutation
relation $[\hat{x},\hat{p}]=i$ (with $\hbar=1$). In analogy to the
qubit Pauli operators, the single mode Pauli operators are defined
as $X(s)=exp[-is\hat{p}]$ and $Z(t)=exp[it\hat{x}]$ with $s,t\in
\mathbb{R}$. The Pauli operator $X(s)$ is a position-translation
operator, which acts on the computational basis of position
eigenstates $\{|q\rangle; q\in \mathbb{R}\}$ as
$X(s)|q\rangle=|q+s\rangle$, whereas $Z$ is a momentum-translation
operator, which acts on the momentum eigenstates as
$Z(t)|p\rangle=|p+t\rangle$. These operators are non-commutative and
obey the identity $ X(s)Z(t)=e^{-ist}Z(t)X(s)$. The Pauli operators
for one mode can be used to construct a set of Pauli operators
$\{X_{i}(s_{i}),Z_{i}(t_{i}); i=1,...,n\}$ for n-mode systems. This
set generates the Pauli group $\mathcal{C}_{1} $. The clifford group
$\mathcal{C}_{2} $ is the normalizer of the Pauli group, whose
transformations acting by conjugating, preserve the Pauli group
$\mathcal{C}_{1} $; i.e., a gate $\emph{U}$ is in the Clifford group
if $\emph{UR}\emph{U}^{-1}\in\mathcal{C}_{1}$ for every
$\emph{R}\in\mathcal{C}_{1}$. The clifford group $\mathcal{C}_{2} $
for CV is shown \cite{seventeen} to be the (semidirect) product of
the Pauli group and linear symplectic group of all one-mode and
two-mode squeezing transformations. Transformation between the
position and momentum basis is given by the Fourier transform
operator $F=exp[i(\pi/4)(\hat{x}^{2}+\hat{p}^{2})]$, with
$F|q\rangle_{x}=|q\rangle_{p}$. The action $FRF^{-1}$ of the Fourier
transform on the Pauli operators is
\begin{eqnarray}
F:X(s)&\rightarrow& Z(s)
,\nonumber \\
Z(t)&\rightarrow& X(-t).  \label{Four}
\end{eqnarray}
This is the generalization of the Hadamard gate for qubits. The
phase gate $ P(\eta)=exp[i(\eta/2)\hat{x}^{2}]$ with $\eta\in
\mathbb{R}$ is a squeezing operation for CV and the action on the
Pauli operators is
\begin{eqnarray}
P(\eta):X(s)&\rightarrow& e^{-is^{2}\eta/2}Z(s\eta)X(s)
,\nonumber \\
Z(t)&\rightarrow& Z(t),  \label{phase}
\end{eqnarray}
in analogy to the phase gate of qubit \cite{eighteen}. The
controlled operation C-Z is generalized to controlled-$ Z (C_{Z})$.
This gate $C_{Z}=exp[i\hat{x}_{1}\bigotimes\hat{x}_{2}]$ provides
the basic interaction for two mode 1 and 2, and describes the
quantum nondemolition (QND) interaction. This set $ \{X(s), F,
P(\eta), C-Z; s,\eta \in \mathbb{R}\}$ generates the Clifford group.
Transformations in the Clifford group do not form a universal set of
gates for CV quantum computation. However, Clifford group
transformation (Gaussian transformations) together with any
higher-order nonlinear transformation (non-Gaussian transformation)
acting on a single-mode form a universal set of gates
\cite{seventeen}. The local Gaussian group only was concerned here,
which can be obtained by repeated application of Fourier and phase
gates. In the following, another type of the phase gate will be used
$ P_{X}(\eta)=FP(\eta)F^{-1}=exp[i(\eta/2)\hat{p}^{2}]$ and the
action on the Pauli operators is
\begin{eqnarray}
P_{X}(\eta):X(s)&\rightarrow& X(s)
,\nonumber \\
Z(t)&\rightarrow& e^{-it^{2}\eta/2}X(-t\eta)Z(t),  \label{phase1}
\end{eqnarray}
where $P_{X}(\eta)^{\dagger}=P_{X}(\eta)^{-1}=P_{X}(-\eta)$.

A graph quantum state is described by a mathematical graph
$G=(V,E)$, i.e. a finite set of $n$ vertices $V$ connected by a set
of edges $E$ \cite{ninteen}.  An $\left\{ a,c\right\} $-path is a order list of vertices $%
a=a_1,a_2,\ldots ,a_{n-1},a_n=c$, such that for all $i$, $a_i$ and
$a_{i+1}$
are adjacent. A connected graph is a graph that has an $\left\{ a,c\right\} $%
-path for any two $a,c\in V$. Otherwise it is referred to as
disconnected. The neighborhood $N_a\subset V$ is defined as the set
of vertices $b$ for which $\left\{ a,b\right\} \in E$. When a vertex
a is deleted in a graph G, together with all edges incident with a,
one obtains a new graph, denoted by $G-a$. For a subset of vertices
$U\subset V$ of a graph $G=(V,E)$ let us denote with $G-U$ the graph
that is obtained from $G$ by deleting the set $U$ of vertices and
all edges which are incident with an element of $U$. Similarly, an
subgraph $G[C]$ of a graph $G=(V,E)$, where $C\subset V$, is
obtained by deleting all vertices and the incident edges that are
not contained in $C$. The preparation procedure of CV graph states
\cite{thirteen} can exactly mirror that for qubit graph states only
using the Clifford operations: first, prepare each mode (or graph
vertex) in a phase-squeezed state, approximating a zero-phase
eigenstate (analog of Pauli-X eigenstates), then, apply a QND
interaction (C-Z gate) to each pair of modes $(j,k)$ linked by an
edge in the graph. All C-Z gates commute. Thus, the resulting CV
graph state becomes, in the limit of infinite squeezing,
$g_{a}=(\hat{p}_{a}-\sum_{b\in N_{a}}\hat{x}_{b})\rightarrow0$,
where the modes $a\in V$ correspond to the vertices of the graph of
$n$ modes, while the modes $b\in N_{a}$ are the nearest neighbors of
mode $a$. This relation is as a simultaneous zero-eigenstate of the
position-momentum linear combination operators. The stabilizers
$G_{a}(\xi)=exp[-i \xi g_{a}]=X_{a}(\xi)\prod_{b\in
N_{a}}Z_{b}(\xi)$ with $\xi\in \mathbb{R}$ for CV graph states are
analogous to $n$ independent stabilizers $G_{a}=X_{a}\prod_{b\in
N_{a}}Z_{b}$ for qubit graph states. Note that the CV graph states
that is discussed here are unweighted since the QND interactions all
have the same strength. For the CV weighted graph states generated
by the different QND interaction strength, the stabilizers become
$G_{a}(\xi)=X_{a}(\xi)\prod_{b\in N_{a}}Z_{b}(\Omega_{ab}\xi)$,
where $\Omega_{ab}$ is the interaction strength between mode a and
b. The CV weighted graph states are more complex, which is not
considered in this paper.

%
\begin{figure}
\centerline{
\includegraphics[width=3in]{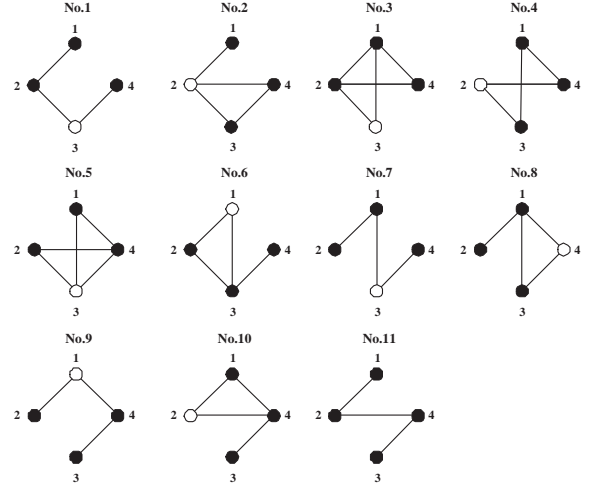}
} \vspace{0.1in}
\caption{ The connected four-vertex graphs for an successive
application of the local complementation. The rule is successively
applied to the vertex, which is circle in the figure. \label{Fig1} }
\end{figure}

The action of the local complement rule, can be described as:
letting $G=(V,E)$ be a graph and $a\in V$ be a vertex, the local
complement of $G$ for $a$, denoted by $\lambda_{a}(G)$, is obtained
by complementing the subgraph of $G$ generated by the neighborhood
$N_{a}$ of $a$ and leaving the rest of the graph unchanged. The
successive application of this rule suffices to generate the
complete orbit of any graph. Here, the corresponding local Gaussian
unitary for CV four-mode graph state were examined. The
corresponding four-mode graph state $|\lambda_{a}(G)\rangle$ by
local complement of a graph $G$ at some vertex $a\in V$, is given by
a local Gaussian unitary operation
\begin{eqnarray}
|\lambda_{a}(G)\rangle=U_{\lambda_{a}}|G\rangle, \label{LC-graph}
\end{eqnarray}
where $U_{\lambda_{a}}$ is local Gaussian operation. A form of the
local Gaussian unitary comprising two types of phase gate is defined
\begin{eqnarray}
U_{LG_{a}}=P_{Xa}(1)\prod_{b\in N_{a}}P_{b}(-1),\label{LC}
\end{eqnarray}
which mirrors the form of qubit local Clifford operation for local
complementation. Fig.1 depicts connected four-mode graphs by such a
successive application of the local complement rule. The four
independent stabilizers of the first graph state $|G^{(1)}\rangle$
are given by
\begin{eqnarray}
G_{1}^{(1)}(\xi)&=&X_{1}(\xi)Z_{2}(\xi),\nonumber \\
G_{2}^{(1)}(\xi)&=&X_{2}(\xi)Z_{1}(\xi)Z_{3}(\xi),\nonumber
\\
G_{3}^{(1)}(\xi)&=&X_{3}(\xi)Z_{2}(\xi)Z_{4}(\xi),\nonumber
\\
G_{4}^{(1)}(\xi)&=&X_{4}(\xi)Z_{3}(\xi).\label{stab1}
\end{eqnarray}
with $G_{i}^{(1)}(\xi)|G^{(1)}\rangle=|G^{(1)}\rangle$ in the limit
of infinite squeezing, where $i=1,...,4$. Applying the local
Gaussian unitary $U_{LG_{3}}$ to the vertex 3, I can compute the
four independent stabilizers of the resulting graph state
$|G^{(2)}\rangle$ by Eqs.
(\ref{phase},\ref{phase1},\ref{LC-graph},\ref{LC}), for example
calculating $G_{2}^{(2)}(\xi)$,
\begin{eqnarray}
|G^{(2)}\rangle&=&|\lambda_{3}(G^{(1)})\rangle\nonumber \\
&=&U_{LG_{3}}G_{2}^{(1)}(\xi)|G^{(1)}\rangle\nonumber \\
&=&U_{LG_{3}}G_{2}^{(1)}(\xi)U_{LG_{3}}^{-1}U_{LG_{3}}|G^{(1)}\rangle \nonumber \\
&=&[e^{i\xi^{2}/2}Z_{2}(-\xi)X_{2}(\xi)]Z_{1}(\xi)\nonumber\\&&
[e^{-i\xi^{2}/2}X_{3}(-\xi)Z_{3}(\xi)]U_{LG_{3}}|G^{(1)}\rangle\nonumber \\
&=&X_{2}(\xi)Z_{1}(\xi)Z_{3}(\xi)U_{LG_{3}}\nonumber\\&&
[Z_{2}(-\xi)X_{3}(-\xi)]|G^{(1)}\rangle\nonumber \\
&=&X_{2}(\xi)Z_{1}(\xi)Z_{3}(\xi)U_{LG_{3}}\nonumber\\&&
[Z_{2}(-\xi)X_{3}(-\xi)]G_{3}^{(1)}(\xi)|G^{(1)}\rangle\nonumber \\
&=&X_{2}(\xi)Z_{1}(\xi)Z_{3}(\xi)Z_{4}(\xi)|\lambda_{3}(G^{(1)})\rangle\nonumber \\
&=&G_{2}^{(2)}(\xi)|G^{(2)}\rangle
\end{eqnarray}
to obtain
\begin{eqnarray}
G_{1}^{(2)}(\xi)&=&X_{1}(\xi)Z_{2}(\xi),\nonumber \\
G_{2}^{(2)}(\xi)&=&X_{2}(\xi)Z_{1}(\xi)Z_{3}(\xi)Z_{4}(\xi),\nonumber
\\
G_{3}^{(2)}(\xi)&=&X_{3}(\xi)Z_{2}(\xi)Z_{4}(\xi),\nonumber
\\
G_{4}^{(2)}(\xi)&=&X_{4}(\xi)Z_{2}(\xi)Z_{3}(\xi),
\end{eqnarray}
which exactly correspond to the stabilizers of No.2 graph state in
Fig.1. The complete orbit of the first graph can be obtained by
applying the local complement rule repeatedly to the vertices and
the corresponding local Gaussian unitary is shown in the following
forms: $No.1\stackrel{U_{LG_{3}}}{\longrightarrow
}No.2\stackrel{U_{LG_{3}}^{2}F^{2}_{1}U_{LG_{2}}^{\dagger}}{\longrightarrow
}No.3\stackrel{U_{LG_{3}}^{\dagger}}{\longrightarrow
}No.4\stackrel{U_{LG_{1}}}{\longrightarrow
}No.5\stackrel{U_{LG_{2}}^{2}F^{2}_{1}U_{LG_{3}}^{\dagger}}{\longrightarrow
}No.6\stackrel{U_{LG_{1}}^{\dagger}}{\longrightarrow
}No.7\stackrel{U_{LG_{3}}}{\longrightarrow
}No.8\stackrel{U_{LG_{4}}^{\dagger}}{\longrightarrow
}No.9\stackrel{U_{LG_{1}}}{\longrightarrow
}No.10\stackrel{U_{LG_{2}}^{\dagger}}{\longrightarrow }No.11$. Here
the complete orbit means the local complement rule is applied on the
graph until exhaust all possibilities.
%
\begin{figure}
\centerline{
\includegraphics[width=3in]{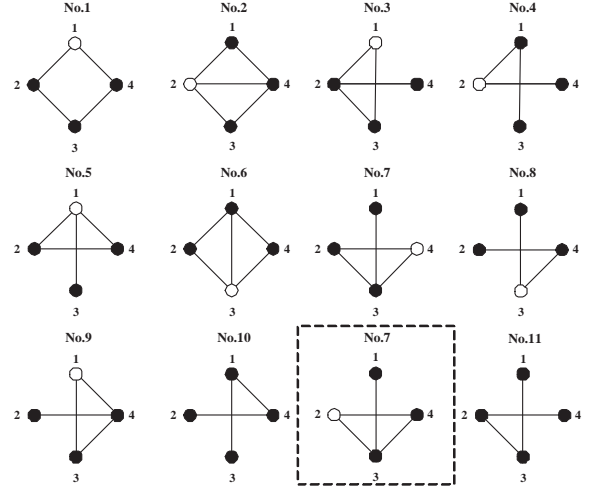}
} \vspace{0.1in}
\caption{ The set of four-vertex graphs is equivalent to Fig.1 under
local Gaussian transformation and graph isomorphisms. The graph
No.7, which is repeated and placed in the dash-line box behind the
No.10, is used for generating the graph No.11 directly. \label{Fig2}
}
\end{figure}
Notice the difference in the Gaussian operations of $2\rightarrow3$,
and $5\rightarrow6$. In the qubit case, these would have been of
identical form. This shows the added richness of CV graph states
over their qubit counterparts. Note that Hein et al. \cite{one}
classify the equivalence of the graph states by considering the
local complementation and additional graph isomorphisms, which
corresponds to the permutations of the vertices. Fig.2 shows another
set of graphs, which are not equivalent to any graph in the
equivalence class represented in Fig.1 only considering the local
complementation. However, they belong to the same equivalence class
when considering both, local Gaussian unitary and graph
isomorphisms. The corresponding local Gaussian unitary in Fig.2 is
shown in the following forms:
$No.1\stackrel{U_{LG_{1}}}{\longrightarrow
}No.2\stackrel{U_{LG_{3}}^{2}F^{2}_{1}U_{LG_{2}}^{\dagger}}{\longrightarrow
}No.3\stackrel{U_{LG_{1}}^{\dagger}}{\longrightarrow
}No.4\stackrel{U_{LG_{2}}}{\longrightarrow
}No.5\stackrel{U_{LG_{2}}^{2}F^{2}_{3}U_{LG_{1}}^{\dagger}}{\longrightarrow
}No.6\stackrel{U_{LG_{4}}^{2}F^{2}_{2}U_{LG_{3}}^{\dagger}}{\longrightarrow
}No.7\stackrel{U_{LG_{4}}^{\dagger}}{\longrightarrow
}No.8\stackrel{U_{LG_{3}}}{\longrightarrow
}No.9\stackrel{U_{LG_{1}}^{\dagger}}{\longrightarrow }No.10;$
 $No.7\stackrel{U_{LG_{2}}^{\dagger}}{\longrightarrow }No.11$.

%
\begin{figure}
\centerline{
\includegraphics[width=3in]{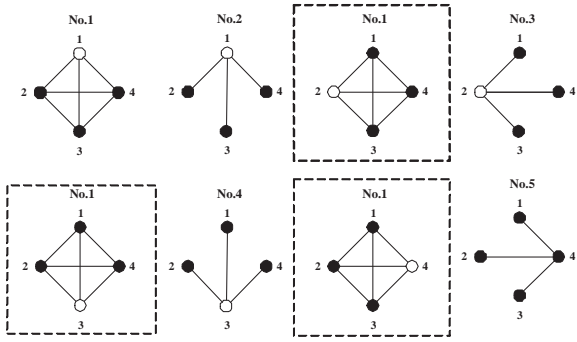}
} \vspace{0.1in}
\caption{ The set of four-vertex graphs is not equivalent to Fig.1
and 2 under local Gaussian transformation and graph isomorphisms.
The graph No.1, which is placed in the dash-line box, is used
repeatedly by the local complementation. \label{Fig3} }
\end{figure}
The set of graphs in Fig.3, usually called GHZ
(Greenberger-Horne-Zeilinger) entangled states, is not equivalent
with Fig.1 and 2 under local Gaussian transformation and graph
isomorphisms. The local Gaussian unitary is applied to four-mode
graph states in Fig.3, which is written above the arrows of the
following diagram:
$No.1\stackrel{U_{LG_{1}}^{\dagger}}{\longrightarrow
}No.2\stackrel{U_{LG_{1}}}{\longrightarrow
}No.1\stackrel{U_{LG_{2}}^{\dagger}}{\longrightarrow
}No.3\stackrel{U_{LG_{2}}}{\longrightarrow
}No.1\stackrel{U_{LG_{3}}^{\dagger}}{\longrightarrow
}No.4\stackrel{U_{LG_{3}}}{\longrightarrow
}No.1\stackrel{U_{LG_{4}}^{\dagger}}{\longrightarrow }No.5$. Fig. 4
lists the graphs with up to four vertices that are not equivalent
under local Gaussian transformation and graph isomorphisms.
%
\begin{figure}
\centerline{
\includegraphics[width=3in]{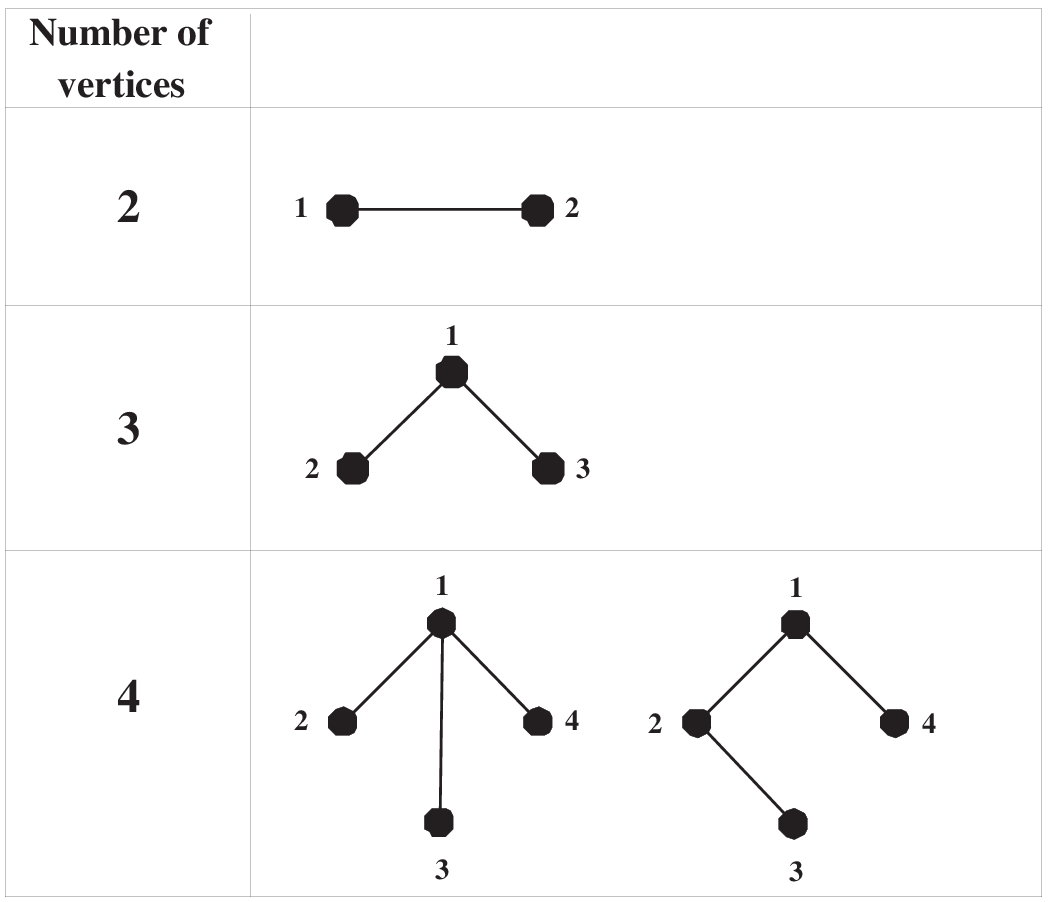}
} \vspace{0.1in}
\caption{ The connected graphs with up to four vertices are not
equivalent under local Gaussian transformation and graph
isomorphisms. \label{Fig4} }
\end{figure}

In summary, the local complement rule was extended for CV graph
states and the corresponding local Gaussian transformations of
four-mode unweighted graph states were given. Thus the local
Gaussian equivalence classes of CV four-mode unweighted graph states
can be obtained. It was shown that the corresponding local Clifford
unitary can not exactly mirror that for qubit and demonstrate the
complexity of CV quantum systems. It is worth remarking that,
whether the local complementation for any CV graph states can be
implemented completely by the local Gaussian transformations and the
general form of the corresponding local Gaussian unitary can be
found, still need be further investigated. This work not only
contribute to a deeper and more complete understanding of CV
multipartite entanglement, but also stimulate the research on CV
graph states theoretically and experimentally.

$^{\dagger} $Corresponding author's email address:
jzhang74@sxu.edu.cn, jzhang74@yahoo.com

\section{\textbf{ACKNOWLEDGMENTS}}

J. Zhang thanks K. Peng and C. Xie for the helpful discussions. This
research was supported in part by NSFC for Distinguished Young
Scholars (Grant No. 10725416), National Basic Research Program of
China (Grant No. 2006CB921101), NSFC (Grant No. 60678029), Program
for the Top Young and Middle-aged Innovative Talents of Higher
Learning Institutions of Shanxi and NSF of Shanxi Province (Grant
No. 2006011003).

\end{document}